\documentclass[conference]{IEEEtran}
\IEEEoverridecommandlockouts
\usepackage{cite}
\usepackage{hyperref}
\usepackage{orcidlink}
\usepackage{amsmath,amssymb,amsfonts}
\usepackage{algorithmic}
\usepackage{booktabs}
\usepackage{graphicx}
\usepackage{textcomp}
\usepackage{xcolor}
\usepackage{subcaption}
\def\BibTeX{{\rm B\kern-.05em{\sc i\kern-.025em b}\kern-.08em
    T\kern-.1667em\lower.7ex\hbox{E}\kern-.125emX}}
\begin{document}

\title{Uncertainty-Aware Ordinal Deep Learning for cross-Dataset Diabetic Retinopathy Grading\\

\thanks{}
}

\author{\IEEEauthorblockN{Ali El Bellaj\orcidlink{0009-0001-4017-2849}\IEEEauthorrefmark{1}\IEEEauthorrefmark{3} \ \  Aya Benradi\orcidlink{0009-0003-5297-4819}\IEEEauthorrefmark{2}\IEEEauthorrefmark{3} \\ Salman El Youssoufi\IEEEauthorrefmark{3}  \ \  Taha El Marzouki\IEEEauthorrefmark{3} \ \   Mohammed-Amine Cheddadi\IEEEauthorrefmark{1}\IEEEauthorrefmark{3}}

\\ 
\IEEEauthorrefmark{1}Department of Computer Science \& Engineering, Mississippi State University, MS, USA \\
\IEEEauthorrefmark{2}Department of Mechanical Engineering, Mississippi State University, MS, USA \\

\IEEEauthorrefmark{3} College of Engineering \& Architecture, International University of Rabat,, Rabat, Morocco
}

\maketitle

\begin{abstract}
Diabetes mellitus is a chronic metabolic disorder characterized by persistent hyperglycemia due to insufficient insulin production or impaired insulin utilization. One of its most severe complications is diabetic retinopathy (DR), a progressive retinal disease caused by microvascular damage, leading to hemorrhages, exudates, and potential vision loss. Early and reliable detection of DR is therefore critical for preventing irreversible blindness. 

In this work, we propose an uncertainty-aware deep learning framework for automated DR severity grading that explicitly models the ordinal nature of disease progression. Our approach combines a convolutional backbone with lesion-query attention pooling and an evidential Dirichlet-based ordinal regression head, enabling both accurate severity prediction and principled estimation of predictive uncertainty. The model is trained using an ordinal evidential loss with annealed regularization to encourage calibrated confidence under domain shift.

We evaluate the proposed method on a multi-domain training setup combining APTOS, Messidor-2, and a subset of EyePACS fundus datasets. Experimental results demonstrate strong cross-dataset generalization, achieving competitive classification accuracy and high quadratic weighted kappa on held-out test sets, while providing meaningful uncertainty estimates for low-confidence cases. These results suggest that ordinal evidential learning is a promising direction for robust and clinically reliable diabetic retinopathy grading.
\end{abstract}


\section{Introduction}

Diabetic retinopathy (DR) is one of the leading causes of preventable blindness worldwide, resulting from long-term microvascular damage to the retina induced by chronic hyperglycemia. As the disease progresses, retinal lesions such as microaneurysms, hemorrhages, and exudates become increasingly prevalent, ultimately threatening vision if left untreated. Regular retinal screening enables early intervention; however, large-scale screening programs remain constrained by limited availability of expert graders and significant inter-observer variability.

Recent advances in deep learning have enabled automated DR detection and grading from fundus photographs with performance approaching that of human experts. Nevertheless, two critical challenges remain. First, DR severity levels are inherently ordinal: disease stages follow a natural progression where misclassifying adjacent grades is less severe than confusing distant ones. Standard classification objectives fail to explicitly encode this ordinal structure. Second, medical decision-making demands not only accurate predictions but also reliable estimates of model confidence, particularly under domain shift across imaging devices, acquisition protocols, and patient populations.

\begin{figure*}
    \centering
    \includegraphics[width=1\linewidth]{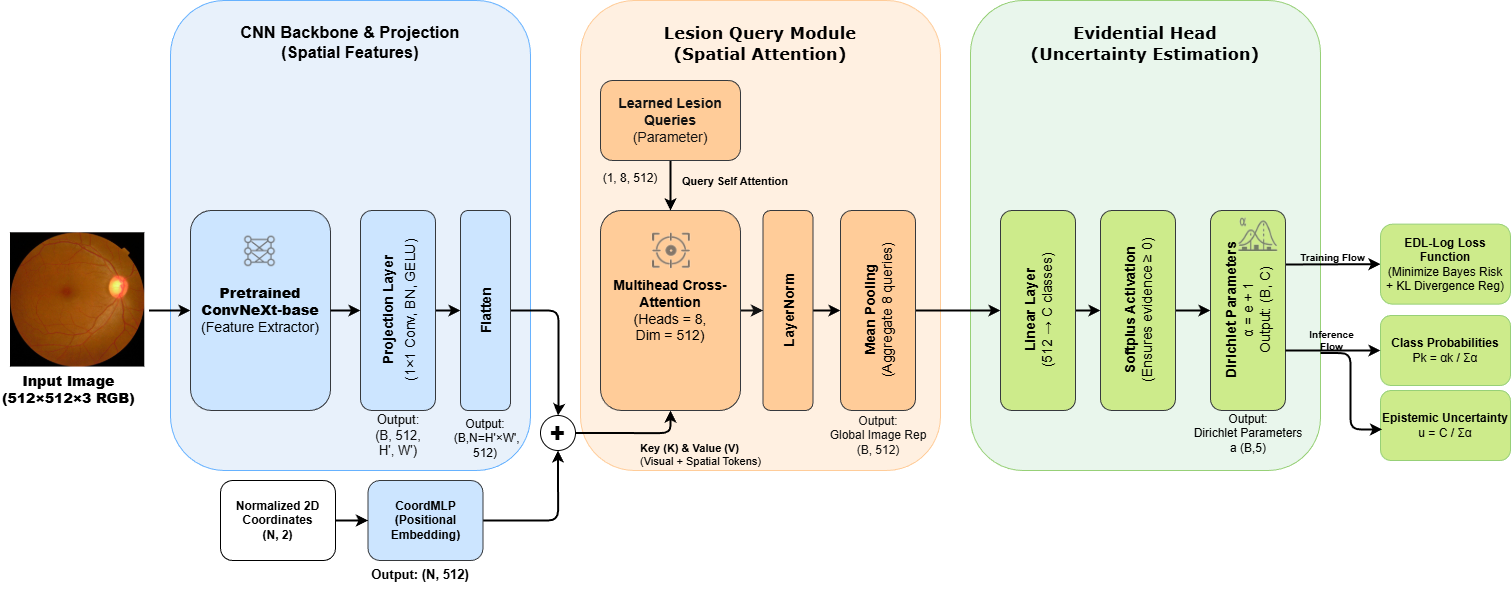}
    \caption{Overview of the proposed lesion-aware ordinal evidential architecture. A pretrained ConvNeXt-Base backbone extracts spatial feature maps that are projected into token embeddings and augmented with positional signals. Learnable lesion queries perform cross-attention pooling to selectively aggregate discriminative retinal regions, producing global representation. An ordinal evidential head then estimates Dirichlet parameters to jointly predict class probabilities and epistemic uncertainty.}
    \label{fig:figure1}
\end{figure*}

To address these limitations, we propose an uncertainty-aware ordinal deep learning framework for DR severity grading. Our method combines a convolutional visual backbone with a lesion-query attention pooling mechanism that aggregates localized pathological features into a global representation. On top of this representation, we introduce an evidential Dirichlet-based ordinal regression head that jointly predicts disease severity and estimates predictive uncertainty. An ordinal evidential loss is employed to respect disease stage ordering while encouraging calibrated confidence through annealed regularization. Figure~\ref{fig:figure1} gives a visual representation of our architecture.

We evaluate the proposed approach in a cross-dataset setting by training on a combination of APTOS \cite{aptos2019-blindness-detection}, Messidor-2 \cite{Decenciere2014Messidor}, and a subset of EyePACS \cite{diabetic-retinopathy-detection} fundus datasets. Our results demonstrate strong generalization across domains, competitive grading performance, we report an accuracy of 87.6\% and a kappa score of 94\%, and meaningful uncertainty estimates that correlate with ambiguous or low-quality images. These findings suggest that combining ordinal learning with evidential uncertainty modeling is a promising direction toward robust and clinically trustworthy DR screening systems.

The main contributions of this work are:
\begin{itemize}
    \item An ordinal evidential learning framework for DR grading that jointly models severity progression and predictive uncertainty.
    \item A lesion-query attention pooling mechanism for aggregating fine-grained retinal lesion information.
    \item A cross-dataset evaluation demonstrating robust generalization across heterogeneous DR benchmarks.
\end{itemize}

\section{Related Work}

\subsection{Deep Learning for Diabetic Retinopathy Grading}

Early applications of convolutional neural networks to DR screening demonstrated that deep models can achieve expert-level performance in detecting referable DR from fundus images \cite{early, early2}. Subsequent work extended these approaches to multi-class DR severity grading using deeper architectures, attention mechanisms, and ensemble strategies \cite{attention1, attention2}. More recent studies have explored transformer-based models and multi-scale feature extraction to capture both global retinal context and small lesion patterns \cite{transformer}. Despite these advances, most existing methods formulate DR grading as a standard multi-class classification task, neglecting the ordinal nature of disease progression.

\subsection{Ordinal Learning in Medical Image Analysis}

Ordinal regression techniques explicitly model ordered class relationships and have been successfully applied to disease staging tasks, including DR grading \cite{ordinal, ABRAHAM2026108614}. Methods such as CORAL and its variants enforce monotonic decision boundaries across severity thresholds, yielding improved consistency compared to flat classification losses. Ordinal learning has also been combined with deep architectures to better reflect clinical progression patterns \cite{ABRAHAM2026108614}. However, these approaches typically provide deterministic predictions and do not quantify predictive uncertainty, which is essential in safety-critical medical applications.

\subsection{Uncertainty Estimation and Evidential Deep Learning}

Quantifying uncertainty in deep neural networks has gained increasing attention for medical imaging applications. Bayesian neural networks, Monte Carlo dropout, and deep ensembles have been explored to estimate epistemic uncertainty \cite{uncertainty, montecarlo}. More recently, evidential deep learning has emerged as an alternative framework that models predictive distributions via Dirichlet evidence, enabling joint estimation of class probabilities and uncertainty without sampling-based inference. Evidential methods have been applied to medical diagnosis and out-of-distribution detection \cite{evidential}, yet their integration with ordinal regression remains relatively underexplored.

\subsection{Cross-Dataset Generalization in Fundus Imaging}

A persistent challenge in DR screening models is generalization across datasets collected under different imaging conditions and demographic distributions. Domain shift between popular benchmarks such as EyePACS, APTOS, and Messidor often leads to performance degradation when models are deployed outside their training distribution. Various strategies, including data augmentation, domain adaptation, and multi-dataset training, have been proposed to mitigate this issue \cite{HACISOFTAOGLU2020409}. However, most prior work focuses on improving accuracy, with limited attention to uncertainty-aware predictions under domain shift.

\subsection{Positioning of This Work}

In contrast to prior approaches, our method unifies ordinal regression and evidential uncertainty modeling within a single end-to-end trainable architecture for DR grading. By incorporating lesion-query attention pooling, we further enhance sensitivity to localized pathological patterns while maintaining global retinal context. To the best of our knowledge, this is among the first works to combine ordinal evidential learning with cross-dataset DR training, explicitly targeting both robust performance and calibrated uncertainty in real-world screening scenarios.

\section{Methodology}

\subsection{Data Preprocessing and Quality Control}

Data preprocessing is a critical precursor to accurate Diabetic Retinopathy grading, as fundus images are frequently characterized by significant noise, varying lighting conditions, and artifacts. To ensure the integrity of the feature extraction process, we implement a two-stage quality control pipeline to eliminate substandard samples.

\subsubsection{Quality Filtration}
First, we address global illumination issues. Images with insufficient lighting often obscure pathological markers such as microaneurysms. We calculate the mean brightness $B_{\mu}$ of each image across the $N$ pixels in the RGB channels:
\begin{equation}
B_{\mu} = \frac{1}{3N} \sum_{i=1}^{N} (R_i + G_i + B_i)
\end{equation}
Images falling below a threshold of $\tau_B = 15$ on a $[0, 255]$ scale are discarded as severely underexposed.

Second, we mitigate the impact of motion blur and focus errors using the Laplacian Variance method. This technique measures the amount of high-frequency detail in an image; a low variance indicates a lack of sharp edges, characteristic of blurriness. We convolve the image with the Laplacian operator $L$:
\begin{equation}
L(x,y) = \frac{\partial^2 I}{\partial x^2} + \frac{\partial^2 I}{\partial y^2}
\end{equation}
The focus score is defined as the variance of the resulting response:
\begin{equation}
\text{Score} = \sigma^2(L(I))
\end{equation}
Images with a variance below our empirically determined threshold are excluded from the training set.

\subsubsection{Normalization and Augmentation Strategy}
While techniques such as Gabor filters and Contrast Limited Adaptive Histogram Equalization (CLAHE) are prevalent in the literature \cite{gabor2025}, \cite{CLAHE}, our experiments indicated that Fundus Cropping provided superior performance by removing non-retinal black borders and maximizing the receptive field of the ConvNeXt backbone on relevant tissue.

In our framework, CLAHE and other geometric transformations are utilized during the training phase as stochastic augmentations rather than static preprocessing. This approach prevents the model from overfitting to specific contrast levels and enhances robustness to domain shifts across datasets such as EyePACS and APTOS. The benefits of each augmentation used in our setup are summarized in Table~\ref{tab:augmentations}.

\begin{table*}[h]
\centering
\caption{Benefits of Data Augmentations in DR Grading}
\label{tab:augmentations}
\begin{tabular}{|l|l|}
\hline
\textbf{Augmentation} & \textbf{Clinical/Technical Benefit} \\ \hline
CLAHE & Enhances visibility of subtle lesions like microaneurysms. \\ \hline
Brightness/Contrast & Mimics variations in camera sensor sensitivity and lighting. \\ \hline
Hue/Saturation & Accounts for differences in retinal pigmentation and dye. \\ \hline
Gaussian Noise/Blur & Improves robustness to image compression and sensor artifacts. \\ \hline
Flips (V/H) & Invariance to eye orientation (Left vs. Right) and patient positioning. \\ \hline
MixUp/CutMix & Improves uncertainty calibration and prevents overconfidence. \\ \hline
\end{tabular}
\end{table*}

\subsection{Backbone and Hierarchical Feature Extraction}

We employ the ConvNeXt-Base architecture as our primary feature extractor. ConvNeXt represents a modern refinement of the classical Convolutional Neural Network (CNN), integrating key design principles from Vision Transformers (ViTs)—such as inverted bottlenecks, large kernel depthwise convolutions, and Layer Normalization—while maintaining the efficiency and inductive biases of pure convolutions.

The architecture is organized into four hierarchical stages, each characterized by a specific spatial resolution and channel depth. For an input image $I \in \mathbb{R}^{512 \times 512 \times 3}$, the initial "patchify" stem downsamples the data into a $128 \times 128$ feature map. Subsequent stages progressively reduce spatial dimensions while enriching the semantic representation:
\begin{itemize}
    \item \textbf{Stage 1:} Produces high-resolution features ($128 \times 128$) primarily capturing low-level textures.
    \item \textbf{Stage 2:} Maintains significant spatial fidelity ($64 \times 64$) while beginning to encode complex anatomical structures and early pathological markers.
    \item \textbf{Stage 3 \& 4:} Provide global semantic context at lower resolutions ($32 \times 32$ and $16 \times 16$), ideal for whole-eye assessment but potentially losing the fine-grained detail of tiny lesions.
\end{itemize}

In our framework, we extract the feature map specifically from Stage 2. This choice is motivated by the clinical requirement to detect microaneurysms and small hemorrhages, which may span only a few pixels. By utilizing Stage 2 as the source for our latent space, we provide the subsequent attention mechanism with a rich, high-resolution feature bank. This enables the model to perform localized query pooling on tokens that retain sufficient spatial information to distinguish pathological nuances before the significant downsampling of later stages.

This hierarchical representation serves as the input to our query-based pooling module, where the model learns to selectively attend to specific retinal regions, as detailed in the following section.

\subsection{Lesion-Query Attention Pooling}

\begin{figure*}[h!]
    \centering
    
    \begin{subfigure}[b]{0.45\textwidth}
        \centering
        \includegraphics[width=\textwidth]{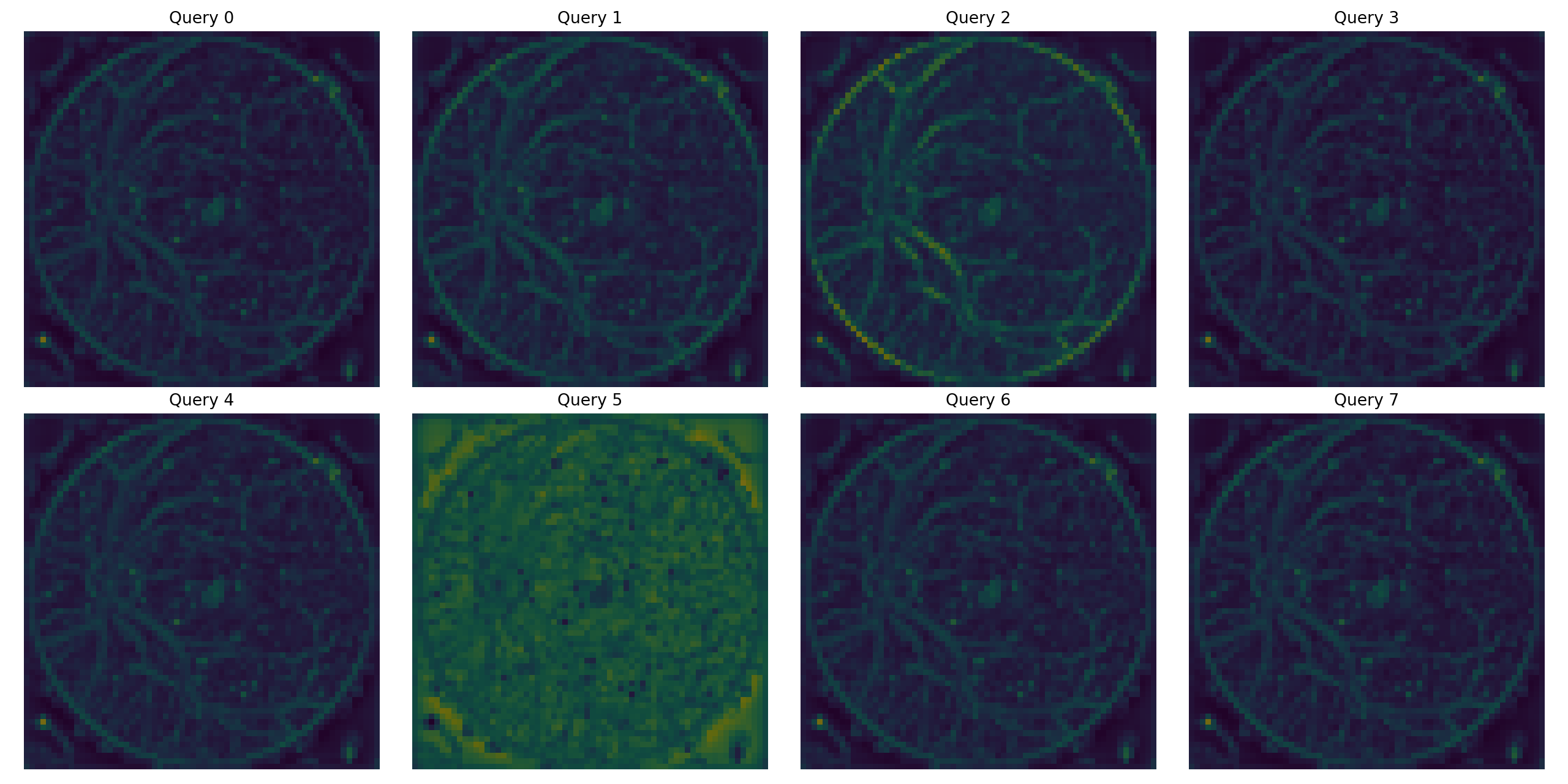} 
        \caption{Query activation maps at \textbf{Epoch 8}. In the early stages of training, queries exhibit broad, overlapping attention regions, primarily responding to global anatomical structures and high-contrast boundaries.}
        \label{fig:left_plot}
    \end{subfigure}
    \hfill 
    \begin{subfigure}[b]{0.45\textwidth}
        \centering
        \includegraphics[width=\textwidth]{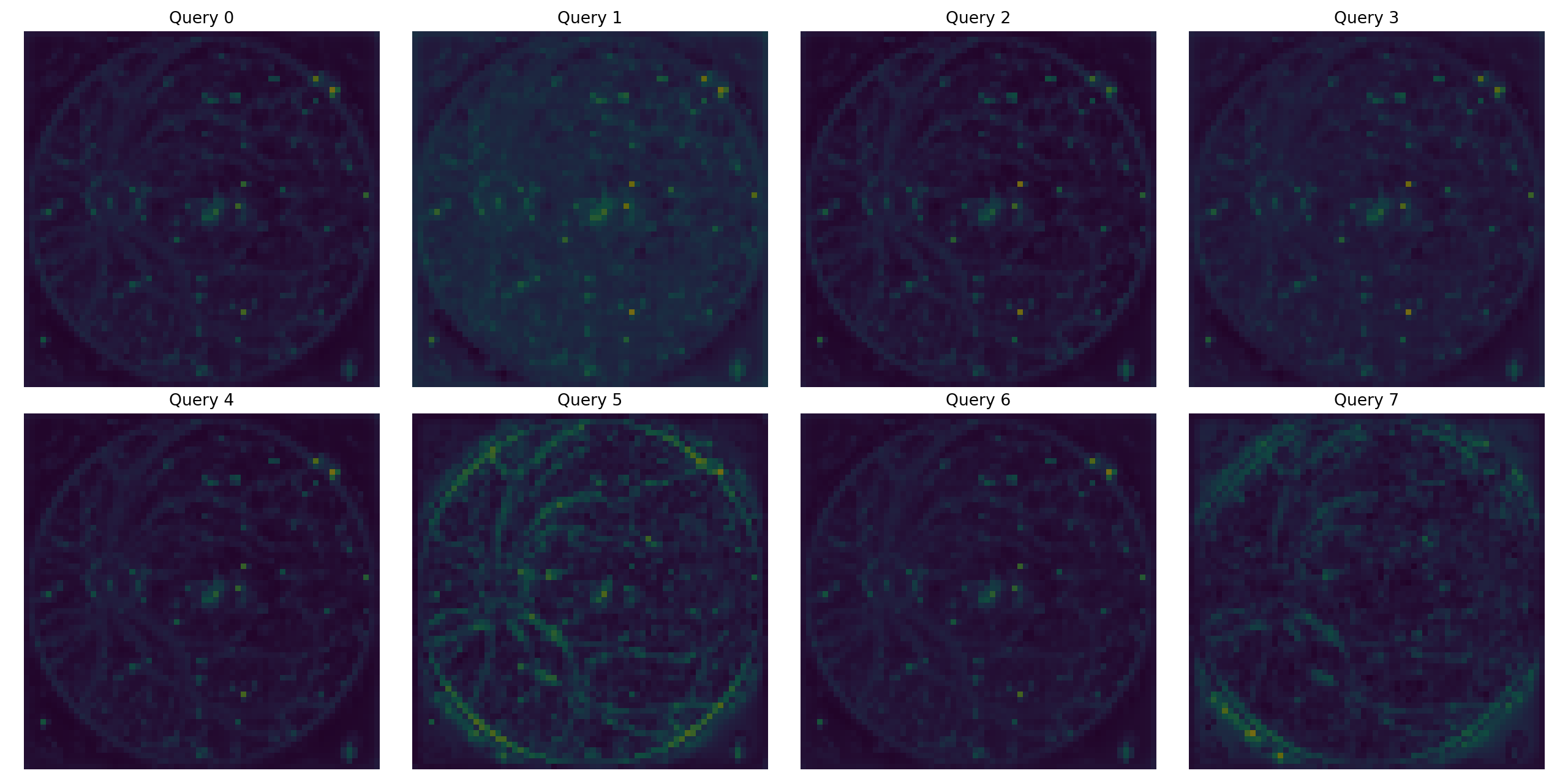}
        \caption{Query activation maps at \textbf{Epoch 19}. As the diversity regularization and temperature scaling take effect, queries demonstrate increased specialization, latching onto distinct, localized clinical indicators with higher spatial precision.}
        \label{fig:right_plot}
    \end{subfigure}
    
    \caption{Evolution of lesion-query specialization during training. (a) Early epoch activations show a lack of differentiation, whereas (b) late epoch activations reveal the model's transition toward independent feature extraction for more robust grading.}
    \label{fig:queries_vis}
\end{figure*}

To bridge the gap between global retinal context and localized pathological markers, we propose a Lesion-Query Attention Pooling (LQAP) module. Unlike standard global average pooling, which collapses spatial information into a single vector, LQAP utilizes a set of learnable latent vectors, or "queries," to selectively aggregate features from the ConvNeXt Stage 2 feature map.

\subsubsection{Transformer Decoder Architecture}
The LQAP module is structured as a multi-layer Transformer decoder, designed to facilitate both feature extraction and inter-query coordination. Each block within the decoder consists of two primary attention mechanisms:

\begin{enumerate}
    \item \textbf{Query Self-Attention:} Before attending to the image, the queries $\mathbf{Q} \in \mathbb{R}^{N \times D}$ undergo self-attention. This allow queries to communicate and coordinate their representations, mitigating the risk of multiple queries collapsing onto the same dominant anatomical feature, such as the optic disc.
    \item \textbf{Cross-Attention:} The queries then interact with the flattened feature tokens $\mathbf{T}$ extracted from the ConvNeXt Stage 2 backbone. The cross-attention mechanism computes a weighted sum of image features based on the similarity between queries and tokens:
    \begin{equation}
    \text{Attention}(\mathbf{Q}, \mathbf{K}, \mathbf{V}) = \text{softmax}\left(\frac{\mathbf{Q}\mathbf{K}^T}{\sqrt{d_k} \cdot \tau}\right)\mathbf{V}
    \end{equation}
    where $\tau$ is a temperature scaling factor used to sharpen the attention distribution, encouraging queries to focus on specific, high-contrast regions.
\end{enumerate}

\subsubsection{Interpretability and Feature Evidence}
The interpretability of the LQAP module is evaluated through query activation maps, as illustrated in Fig.~\ref{fig:queries_vis}. Although the queries do not always exhibit systematic or mutually exclusive specialization for specific lesions, they consistently latch onto localized regions containing clinical indicators such as microaneurysms, hard exudates, or neovascularization. Consequently, these latent queries serve as strong spatial evidence for the model's final ordinal prediction. By mapping these query responses back to the spatial domain, we establish a more transparent reasoning path that informs the subsequent evidential uncertainty head, allowing for a better understanding of which retinal regions contribute most to the predicted severity grade.

\subsection{Evidential Ordinal Head and Training Objective}

\subsubsection{Ordinal thresholds and evidential distribution}
Diabetic retinopathy severity grades follow a natural order, therefore we formulate prediction through $T = K-1$ ordinal thresholds rather than flat $K$-way classification. For class labels $y \in \{0,\dots,K-1\}$, we define threshold events
\begin{equation}
z_k = \mathbb{I}[y > k], \qquad k=0,\dots,T-1,
\end{equation}
where each $z_k \in \{0,1\}$ indicates whether the disease severity exceeds threshold $k$.

Given the pooled latent representation produced by the lesion-query attention pooling (LQAP) module, we predict non-negative \emph{evidence} for each ordinal threshold. For threshold $k$, the network outputs evidence vector $\mathbf{e}_k = [e_{k,0}, e_{k,1}]$, $e_{k,c} \ge 0$. This evidence parameterizes a 2-dimensional Dirichlet distribution (equivalently, a Beta distribution) over the Bernoulli probabilities of $z_k$:
\begin{equation}
\boldsymbol{\alpha}_k = \mathbf{e}_k + \mathbf{1}, \qquad
\mathbf{p}_k \sim \mathrm{Dirichlet}(\boldsymbol{\alpha}_k),
\end{equation}
where $\boldsymbol{\alpha}_k = [\alpha_{k,0}, \alpha_{k,1}]$ are the concentration parameters. The expected probability of exceeding the threshold is then
\begin{equation}
\hat{\pi}_k = \mathbb{E}[p_{k,1}] = \frac{\alpha_{k,1}}{\alpha_{k,0}+\alpha_{k,1}}.
\end{equation}
We recover class probabilities by enforcing ordinal consistency through cumulative differences:
\begin{equation}
\hat{P}(y=0)=1-\hat{\pi}_0,\quad
\end{equation}
\begin{equation}
\hat{P}(y=c)=\hat{\pi}_{c-1}-\hat{\pi}_{c},\quad
\hat{P}(y=K-1)=\hat{\pi}_{T-1}.
\end{equation}

\subsubsection{Ordinal evidential loss with KL annealing}
To train the threshold predictions, we define target thresholds $t_k = \mathbb{I}[y>k]$. When mixup/cutmix is used, we instead use \emph{soft ordinal targets} $t_k \in [0,1]$ computed from the mixed class-probability targets by $t_k=\sum_{c=k+1}^{K-1}P(y=c)$ (soft threshold exceedance).

We optimize an ordinal evidential objective consisting of a data-fit term plus an annealed regularizer:
\begin{equation}
\mathcal{L}_{\text{EDL}} =
\sum_{k=0}^{T-1} \mathcal{L}_{\text{data}}(\hat{\pi}_k, t_k)
\;+\; \lambda(t)\sum_{k=0}^{T-1}\mathrm{KL}\!\left(\mathrm{Dir}(\boldsymbol{\alpha}_k)\,\|\,\mathrm{Dir}(\mathbf{1})\right),
\end{equation}
where $\mathcal{L}_{\text{data}}$ is a binary loss applied per threshold (e.g., BCE on $\hat{\pi}_k$ versus $t_k$). The KL term penalizes unwarranted high evidence by pulling the predictive Dirichlet toward the uniform prior $\mathrm{Dir}(\mathbf{1})$ when the model is uncertain.

To avoid under-confident training early on, we use KL annealing:
\begin{equation}
\lambda(t) = \lambda_{\max}\cdot \min\left(1,\frac{t}{t_{\text{anneal}}}\right),
\end{equation}
so regularization ramps up gradually over the first $t_{\text{anneal}}$ epochs. This improves calibration under domain shift while still allowing the model to discover discriminative lesion cues early in training. The resulting uncertainty proxy can be computed from total evidence, e.g., via the Dirichlet strength $S_k=\alpha_{k,0}+\alpha_{k,1}$ aggregated across thresholds.

\subsubsection{Diversity and specialization penalties for lesion queries}
While query-based pooling provides interpretability, multiple queries may collapse to the same dominant region (e.g., optic disc), reducing coverage of small lesions. We therefore add auxiliary penalties that encourage query specialization and balanced utilization, consistent with the design goal of coordinated queries through self-attention and sharpened cross-attention (Eq.~(4)).

\textbf{(i) Query diversity.} Let $\mathbf{Q}\in\mathbb{R}^{N\times D}$ denote the final query embeddings output by the decoder. We penalize excessive similarity between queries using a cosine-similarity-based term:
\begin{equation}
\mathcal{L}_{\text{div}} = \frac{1}{N(N-1)}\sum_{i\neq j}\left(\max(0,\cos(\mathbf{q}_i,\mathbf{q}_j)-m)\right)^2,
\end{equation}
where $m$ is a margin controlling how orthogonal the queries should be.

\textbf{(ii) Load balancing.} The learned query pooling produces per-image weights $w\in\mathbb{R}^{N}$ indicating which queries contribute to the pooled representation. To prevent only a small subset of queries from dominating, we encourage the batch-average usage to be close to uniform:
\begin{equation}
\mathcal{L}_{\text{lb}} = \left\|\frac{1}{B}\sum_{b=1}^{B}\mathbf{w}^{(b)} - \frac{1}{N}\mathbf{1}\right\|_2^2.
\end{equation}

\textbf{(iii) Spatial entropy range penalty (optional).} To avoid both overly-uniform attention maps (``everything bright'') and single-pixel collapse, we regularize the entropy of each query attention map to stay within a safe range $[h_{\min},h_{\max}]$, optionally weighted by the pooling weight $w$ so that only actively-used queries are constrained.

The full training objective is:
\begin{equation}
\mathcal{L} =
\mathcal{L}_{\text{EDL}}
+ \beta\,\mathcal{L}_{\text{div}}
+ \gamma\,\mathcal{L}_{\text{lb}}
+ \eta\,\mathcal{L}_{\text{sp-ent}},
\end{equation}
where $(\beta,\gamma,\eta)$ control the strength of specialization regularization. In practice, these auxiliary terms improve stability and interpretability of query activation maps by encouraging distinct lesion-focused evidence while preserving ordinal calibration through evidential learning.

\section{Experiments and Results}

\subsection{Experimental Setup}
\textbf{Dataset and split.} We train and evaluate on the diabetic retinopathy (DR) dataset with five ordinal severity grades ($K{=}5$). The dataset is split into train/validation/test subsets following the standard folder structure used in our pipeline.

\textbf{Pre-processing and augmentation.} All images are resized to $512{\times}512$. As pre-processing, we apply fundus cropping followed by a Gaussian smoothing filter and CLAHE to enhance local contrast. During training, we use standard geometric and photometric augmentations (random flips, rotations, brightness/contrast, and mild blur/noise), while validation and testing use deterministic transforms only.

\textbf{Training details.} We train for up to 60 epochs with AdamW, cosine learning rate scheduling with warmup, EMA model averaging, and early stopping based on validation quadratic weighted kappa (QWK). We report final test results using the best checkpoint (by validation QWK), evaluated with EMA and test-time augmentation (TTA) when enabled. We use mixed precision (bf16) when available.

\textbf{Metrics.} We report \emph{Accuracy} and \emph{Quadratic Weighted Kappa (QWK)} as primary metrics, and include \emph{Precision} and \emph{Recall} and averaged (macro average) to reflect class imbalance. We also report uncertainty summary statistics derived from the evidential head (Table~\ref{tab:baseline}).

\subsection{Compared Methods}
We compare the proposed lesion-query evidential ordinal model against the following baselines:

\begin{itemize}
    
    \item \textbf{B1: ConvNeXt backbone + ordinal thresholds (non-evidential).} Threshold-based ordinal head trained with BCE on cumulative exceedance probabilities (no evidential modeling).
    \item \textbf{B2 : ConvNeXt + query pooling (no evidential).} Lesion-query attention pooling (LQAP) followed by an ordinal head, trained without evidential regularization.
    \item \textbf{Ours: ConvNeXt + LQAP + evidential ordinal head.} Our full method: stage-2 token querying, temperature-scaled attention, evidential threshold head with KL annealing, and query specialization regularization.
\end{itemize}

For methods using LQAP, we use $N$ learned queries, query dropout, and (when enabled) diversity/load-balancing penalties to avoid query collapse.

\subsection{Main Results}
Table~\ref{tab:baseline} compares ConvNeXt-based ordinal baselines with and without lesion query attention pooling (LQAP) against the proposed evidential formulation. 
While both non-evidential ordinal baselines achieve competitive quadratic weighted kappa (QWK), incorporating LQAP improves lesion sensitivity and reduces predictive uncertainty. 
Specifically, LQAP lowers the mean uncertainty from $0.166$ to $0.098$, indicating more confident and localized feature aggregation, while maintaining comparable classification performance.

The proposed evidential model further improves performance across all metrics, achieving the highest accuracy (0.876) and QWK (0.940), alongside superior recall and precision. 
Importantly, evidential modeling enables principled uncertainty estimation by explicitly quantifying evidence through Dirichlet concentration parameters, leading to better-calibrated predictions compared to the non-evidential baselines. 

These results demonstrate that combining lesion-focused attention pooling with evidential ordinal learning provides complementary benefits: LQAP enhances spatial specialization and fine-grained lesion detection, while evidential learning improves robustness and calibration. 
Together, they produce both higher diagnostic accuracy and more reliable uncertainty estimates, which are critical for safety-sensitive clinical deployment.

\begin{table}[t]
\centering
\caption{Comparison of ConvNeXt-Base baselines and the proposed method on the test set.}
\label{tab:baseline}
\begin{tabular}{lccccc}
\toprule
Model & Acc & QWK & Recall  & Precision & $u_{\text{mean}}$

\\
\midrule
ConvNeXt (Ordinal) & 0.826 & 0.918 & 0.71 & 0.72 & 0.166 \\
ConvNeXt + LQAP    & 0.823 & 0.914 & 0.72 & 0.73 & 0.098 \\
\textbf{Ours: LQAP + EDL} & \textbf{0.876} & \textbf{0.940} & \textbf{0.75} & \textbf{0.80} & \textbf{0.135} \\
\bottomrule
\end{tabular}
\end{table}

\subsection{Ablation Study}
We perform ablations to quantify the contribution of each component:

\begin{itemize}
    \item \textbf{A1: Stage selection.} Querying stage-2 tokens yields higher QWK than stage-3 querying, consistent with the higher spatial resolution and the small size of DR lesions.
    \item \textbf{A2: Query count and routing.} Using more queries increases capacity but may introduce collapse without regularization. A moderate number of queries with load-balancing and diversity terms yields the most stable training.  Our findings show that 8 queries remain the strongest option.
    \item \textbf{A3: Regularization.} KL annealing improves training stability and prevents premature under-confidence; specialization penalties reduce query redundancy, avoid query collapse and improve interpretability of attention maps.
\end{itemize}

We report an ablation table (Table~\ref{tab:ablation}).

\begin{table}[t]
\centering
\caption{Ablation summary.}
\label{tab:ablation}
\begin{tabular}{lcc}
\hline
\textbf{Variant} & \textbf{Acc} & \textbf{QWK} \\
\hline
Stage-3 querying (EDL) & 0.843 & 0.92 \\
Stage-2 querying (EDL) & 0.876 & 0.94 \\
No load balancing & 0.83 & 0.91 \\
No diversity penalty &  0.83 & 0.90 \\
No KL annealing & 0.82 & 0.90 \\
\hline
\end{tabular}
\end{table}

\subsection{Uncertainty Analysis and Interpretation}
A key advantage of evidential learning is that the model outputs both predictions and uncertainty. Each ordinal threshold outputs Dirichlet parameters $\boldsymbol{\alpha}_k$, whose strength $S_k=\sum_c \alpha_{k,c}$ reflects confidence: low evidence (small $S_k$) indicates uncertainty, while high evidence indicates confidence. We summarize uncertainty per image using an aggregate metric (e.g., mean uncertainty across thresholds).

\textbf{Uncertainty vs. correctness.} We observe that misclassified samples tend to have higher uncertainty than correctly classified samples. This is expected: if the evidence for the correct ordinal thresholds is weak or inconsistent, the Dirichlet strength decreases and uncertainty increases.

\textbf{Failure modes.} High-uncertainty cases often correspond to low-quality images, blur, illumination artifacts, or atypical anatomical structures. These are clinically meaningful: uncertainty can be used to trigger referral to a specialist or request re-acquisition of the image.

\textbf{Practical interpretation.} In a screening workflow, uncertainty can be used as a triage signal: low-uncertainty predictions can be auto-reported, while high-uncertainty cases are flagged for review. This makes the system safer by reducing confident errors.

\subsection{Qualitative Evidence: Query Attention Maps}
We visualize per-query patch attention maps to inspect what each query learns. Early in training, multiple queries respond broadly as seen in Figure~\ref{fig:queries_vis}; later, specialized queries emerge that focus on distinct retinal regions. When collapse occurs (only 1--2 active queries), load balancing and diversity regularization help redistribute query usage. These visualizations support that the proposed architecture learns lesion-relevant evidence rather than relying only on global appearance cues.

\section{Discussion}

The experimental results highlight three main observations regarding lesion-aware pooling, ordinal modeling, and uncertainty estimation.

First, lesion query attention pooling (LQAP) consistently improves spatial selectivity compared to global average pooling. By replacing uniform aggregation with a small set of learnable queries, the model focuses on clinically relevant retinal regions such as microaneurysms, hemorrhages, and exudates.

Second, incorporating the ordinal formulation provides a more appropriate learning objective than standard categorical classification. Diabetic Retinopathy severity is inherently ordered, and modeling predictions through cumulative ordinal thresholds allows the network to exploit inter-class relationships. This reduces extreme misclassifications (e.g., Grade 0 vs. Grade 4), which are heavily penalized in quadratic weighted kappa (QWK). As observed in Table~\ref{tab:baseline}, ordinal modeling consistently increases both recall and kappa compared to non-ordinal baselines.

Third, evidential learning improves both performance and reliability. Instead of producing point probabilities, the evidential head estimates Dirichlet concentration parameters that explicitly encode prediction confidence. This enables the network to express uncertainty under ambiguous or low-quality inputs, resulting in better calibration. The lower mean uncertainty values and improved robustness suggest that evidential modeling discourages overconfident mistakes, a critical requirement for medical decision support systems.

Importantly, these components provide complementary benefits. LQAP enhances spatial localization, ordinal learning exploits label structure, and evidential inference regularizes predictions through uncertainty awareness. When combined, they produce the best overall performance, achieving the highest accuracy and QWK while maintaining well-calibrated confidence estimates.

We also observed that query collapse can occur if regularization is insufficient. Diversity penalties, load-balancing constraints, and temperature scheduling were effective in encouraging queries to specialize across distinct retinal regions. These mechanisms helped preventing multiple queries from attending to identical features and improve representation diversity.

Despite these gains, several limitations remain. The query mechanism introduces additional parameters and computational overhead compared to global pooling. Furthermore, uncertainty estimation currently relies on distributional evidence rather than explicit out-of-distribution detection.

\section{Conclusion}

This paper introduced a lesion-aware ordinal evidential framework for diabetic retinopathy grading that integrates three key ideas: (i) query-based attention pooling for spatial specialization, (ii) ordinal threshold modeling to respect severity ordering, and (iii) evidential learning for principled uncertainty estimation.

Unlike conventional global pooling approaches, the proposed lesion query attention pooling mechanism enables the network to selectively aggregate features from clinically informative regions. Combined with ordinal classification and evidential inference, this leads to both improved predictive accuracy and better-calibrated uncertainty. Experimental results demonstrate consistent gains across all evaluation metrics, achieving state-of-the-art performance in accuracy and quadratic weighted kappa while providing meaningful uncertainty estimates for safety-critical decisions.

These findings suggest that explicitly modeling spatial focus, label structure, and predictive confidence is essential for reliable medical image classification. Beyond diabetic retinopathy, the proposed framework can be readily extended to other ordinal medical grading tasks such as glaucoma staging, cancer severity assessment, or histopathology scoring.

Future work will investigate larger retinal foundation backbones, multi-scale query routing, and improved uncertainty calibration techniques, as well as validation on additional clinical datasets. We believe that combining interpretable attention mechanisms with uncertainty-aware learning constitutes a promising direction toward trustworthy AI-assisted diagnosis.

\bibliographystyle{IEEEtran}
\bibliography{references.bib}

@inproceedings{gabor2025,
author = {Richardy Kurniawan, Adrian and Irmawati, Irmawati},
title = {Detection of Diabetic Retinopathy in Retinal Fundus Images Using Gabor Filters and ConvNeXt},
year = {2025},
isbn = {9798400709586},
publisher = {Association for Computing Machinery},
address = {New York, NY, USA},
url = {https://doi.org/10.1145/3705927.3705936},
doi = {10.1145/3705927.3705936},
booktitle = {Proceedings of the 2024 7th International Conference on Digital Medicine and Image Processing},
pages = {47–52},
numpages = {6},
location = {
},
series = {DMIP '24}
}

@INPROCEEDINGS{early,
  author={Dhope, Prashant and Kulkarni, Meghana},
  booktitle={2025 International Conference on Biomedical Engineering and Sustainable Healthcare (ICBMESH)}, 
  title={Recognition of Diabetic Retinopathy using Deep Learning}, 
  year={2025},
  volume={},
  number={},
  pages={1-5},
  doi={10.1109/ICBMESH66209.2025.11182234}}

@INPROCEEDINGS{early2,
  author={Jiang, Hongyang and Yang, Kang and Gao, Mengdi and Zhang, Dongdong and Ma, He and Qian, Wei},
  booktitle={2019 41st Annual International Conference of the IEEE Engineering in Medicine and Biology Society (EMBC)}, 
  title={An Interpretable Ensemble Deep Learning Model for Diabetic Retinopathy Disease Classification}, 
  year={2019},
  volume={},
  number={},
  pages={2045-2048},
  doi={10.1109/EMBC.2019.8857160}}

@INPROCEEDINGS{attention1,
  author={Hossain, Saima and Hossain, A. B. M. Aowlad},
  booktitle={2024 IEEE International Women in Engineering (WIE) Conference on Electrical and Computer Engineering (WIECON-ECE)}, 
  title={Improving Diabetic Retinopathy Classification by Employing Convolutional Block Attention Mechanism in Modified InceptionV3 Model}, 
  year={2024},
  volume={},
  number={},
  pages={427-431},
  doi={10.1109/WIECON-ECE64149.2024.10914921}}

@INPROCEEDINGS{attention2,
  author={Wu, Keliang and Peng, Jincheng and Feng, Xiang and Chen, Zixuan and Monday, Happy Nkanta and Nneji, Grace Ugochi},
  booktitle={2024 IEEE 16th International Conference on Advanced Infocomm Technology (ICAIT)}, 
  title={Attention-Enhanced Ensemble Learning for Diabetic Retinopathy Classification with Interpretability}, 
  year={2024},
  volume={},
  number={},
  pages={228-233},
  doi={10.1109/ICAIT62580.2024.10807960}}

@INPROCEEDINGS{transformer,
  author={Haque, Saeedul and Samad, Mst Fateha},
  booktitle={2025 2nd International Conference on Next-Generation Computing, IoT and Machine Learning (NCIM)}, 
  title={Lightweight Vision Transformer Network for Diabetic Retinopathy Classification Through Data Balancing Approach}, 
  year={2025},
  volume={},
  number={},
  pages={1-6},
  doi={10.1109/NCIM65934.2025.11160060}}

@INPROCEEDINGS{ordinal,
  author={Lawate, Aditya and Parab, Prachi and Kokate, Onkar and Surti, Neha},
  booktitle={2025 International Conference on Computing and Communication Technologies (ICCCT)}, 
  title={Early Detection of Diabetic Retinopathy Using Deep Learning}, 
  year={2025},
  volume={},
  number={},
  pages={1-6},
  doi={10.1109/ICCCT63501.2025.11020520}}

@article{ABRAHAM2026108614,
title = {MHSA-enhanced CNNs with TOPSIS-driven ensemble learning for automated diabetic retinopathy grading},
journal = {Biomedical Signal Processing and Control},
volume = {112},
pages = {108614},
year = {2026},
issn = {1746-8094},
doi = {https://doi.org/10.1016/j.bspc.2025.108614},
url = {https://www.sciencedirect.com/science/article/pii/S1746809425011255},
author = {Shilpa Elsa Abraham and Binsu C. Kovoor}
}

@ARTICLE{uncertainty,
  author={Jaskari, Joel and Sahlsten, Jaakko and Damoulas, Theodoros and Knoblauch, Jeremias and Särkkä, Simo and Kärkkäinen, Leo and Hietala, Kustaa and Kaski, Kimmo K.},
  journal={IEEE Access}, 
  title={Uncertainty-Aware Deep Learning Methods for Robust Diabetic Retinopathy Classification}, 
  year={2022},
  volume={10},
  number={},
  pages={76669-76681},
  doi={10.1109/ACCESS.2022.3192024}}

@INPROCEEDINGS{montecarlo,
  author={Padmini, B and Kalpana, Y},
  booktitle={2023 International Conference on New Frontiers in Communication, Automation, Management and Security (ICCAMS)}, 
  title={Early detection of DR with an Effective Optimal Stochastic Deep Network in fundus images using the Monte Carlo Method}, 
  year={2023},
  volume={1},
  number={},
  pages={1-8},
  doi={10.1109/ICCAMS60113.2023.10526100}}

@InProceedings{evidential,
author="Wang, Meng
and Wang, Lianyu
and Xu, Xinxing
and Zou, Ke
and Qian, Yiming
and Goh, Rick Siow Mong
and Liu, Yong
and Fu, Huazhu",
editor="Greenspan, Hayit
and Madabhushi, Anant
and Mousavi, Parvin
and Salcudean, Septimiu
and Duncan, James
and Syeda-Mahmood, Tanveer
and Taylor, Russell",
title="Federated Uncertainty-Aware Aggregation for Fundus Diabetic Retinopathy Staging",
booktitle="Medical Image Computing and Computer Assisted Intervention -- MICCAI 2023",
year="2023",
publisher="Springer Nature Switzerland",
address="Cham",
pages="222--232",
isbn="978-3-031-43895-0"
}

@article{HACISOFTAOGLU2020409,
title = {Deep learning frameworks for diabetic retinopathy detection with smartphone-based retinal imaging systems},
journal = {Pattern Recognition Letters},
volume = {135},
pages = {409-417},
year = {2020},
issn = {0167-8655},
doi = {https://doi.org/10.1016/j.patrec.2020.04.009},
url = {https://www.sciencedirect.com/science/article/pii/S016786552030129X},
author = {Recep E. Hacisoftaoglu and Mahmut Karakaya and Ahmed B. Sallam}
}

@INPROCEEDINGS{CLAHE,
  author={Kalita, Deepjyoti and Patoju, Shashank and Dash, Abhipsha and Sharma, Hrishita and Mirza, Khalid B.},
  booktitle={2025 IEEE 6th India Council International Subsections Conference (INDISCON)}, 
  title={Impact of CLAHE and Gaussian Smoothing for Optimized Deep Learning Based Diabetic Retinopathy Detection}, 
  year={2025},
  volume={},
  number={},
  pages={1-6},
  doi={10.1109/INDISCON66021.2025.11252065}}

@misc{aptos2019-blindness-detection,
    author = {Karthik and Maggie and Sohier Dane},
    title = {APTOS 2019 Blindness Detection},
    year = {2019},
    howpublished ={\url{https://kaggle.com/competitions/aptos2019-blindness-detection}},
    note = {Kaggle}
}

@article{Decenciere2014Messidor,
  author  = {Decenci{\`e}re, Etienne and Zhang, Xiwei and Cazuguel, Guy and Lay, Bruno and Cochener, Béatrice and Trone, Charles and Gain, Patrick and Ordóñez, Rosendo and Massin, Pascale and Erginay, Ali and Charton, Beno{\^i}t and Klein, Jean-Claude},
  title   = {Feedback on a publicly distributed image database: the Messidor database},
  journal = {Image Analysis \& Stereology},
  volume  = {33},
  number  = {3},
  pages   = {231--234},
  year    = {2014},
  issn    = {1854-5165}
}

@misc{diabetic-retinopathy-detection,
    author = {Emma Dugas and Jared and Jorge and Will Cukierski},
    title = {Diabetic Retinopathy Detection},
    year = {2015},
    howpublished = {\url{https://kaggle.com/competitions/diabetic-retinopathy-detection}},
    note = {Kaggle}
}

\end{document}